\documentclass[twocolumn,showpacs,pra,floatfix,superscriptaddress]{revtex4-1}
\usepackage{multirow,amssymb,amsbsy,amsmath}
\usepackage{graphicx}
\usepackage{verbatim}
\usepackage[colorlinks=true,linkcolor=blue,citecolor=blue]{hyperref}
\usepackage{subfigure}
\usepackage{braket}

\begin{document}

\title{Local witness for bipartite quantum discord}

\author{Manuel Gessner}

\email{manuel.gessner@physik.uni-freiburg.de}

\affiliation{Physikalisches Institut, Universit\"at Freiburg,
Hermann-Herder-Strasse 3, D-79104 Freiburg, Germany}

\affiliation{Department of Physics, University of California, Berkeley, California 94720, USA}

\author{Heinz-Peter Breuer}

\email{breuer@physik.uni-freiburg.de}

\affiliation{Physikalisches Institut, Universit\"at Freiburg,
Hermann-Herder-Strasse 3, D-79104 Freiburg, Germany}

\date{\today}

\begin{abstract}
Recently, we have proposed a method for the local detection of 
quantum correlations on the basis of local measurements and state tomography at 
different instances in time [Phys.~Rev.~Lett. \textbf{107}, 180402 (2011)]. The 
method allows for the detection of quantum discord in bipartite systems when 
access is restricted to only one of the subsystems. Here, we elaborate the details 
of this method and provide applications to specific physical models. In particular, 
we discuss the performance of the scheme for generic complex systems by 
investigating thermal equilibrium states corresponding to randomly generated 
Hamiltonians. Moreover, we formulate an ergodicity-like hypothesis which links 
the time average to the analytically obtained average over the group of unitary 
operators equipped with the Haar measure.
\end{abstract}


\maketitle
 
\section{Introduction}

The field of quantum information theory is dedicated to developing 
computational techniques with an advantage over classical methods 
using the laws of quantum mechanics \cite{NIELSEN}. A variety of tools for 
communication and computation science have been developed in the past years, 
ranging from quantum teleportation \cite{Bennett93} and quantum dense coding 
\cite{Bennett92} to efficient algorithms for quantum computers 
\cite{Feynman,Shor,KnillLaflamme}. The fundamental resource for these 
applications is usually summarized under the term quantum correlations, even 
though it has proven difficult to identify a common resource to all of these 
applications. More precisely, ideas like quantum teleportation and the violation of 
Bell's inequalities \cite{Bell} are profoundly related to quantum entanglement 
\cite{HHHH}. Other applications could be linked directly to a resource named 
quantum discord \cite{Datta,Lanyon,Dakic}, which is identical to entanglement for 
pure states but differs for statistical mixtures 
\cite{VEDRAL,ZUREK,MODIREVIEW}. While for 
entanglement the term quantum correlation is suitable, not least in 
view of its connection to nonlocality, quantum discord indicates the 
presence of non-commuting local observables in the decomposition of the state 
which does not necessarily imply strong correlations 
\cite{DVB,GUO,BRUSS,CIC,Hu,TURKU}. However, regardless of its interpretation 
in terms of correlations, quantum discord has proven to be an important resource 
for certain tasks in quantum communication and computation \cite{Datta,Lanyon,Dakic,MODIREVIEW}. It is considered 
especially promising in the context of operations involving highly mixed
states, which emerge naturally due to the inevitable influence of noise 
\cite{BREUERBOOK}.

Several methods have been developed which allow for the detection 
of quantum discord with relatively small effort if all subsystems are 
under sufficient degree of control \cite{DVB,ZHANG,AGUILAR,GIROLAMI}. 
Recently, we have shown that the quantum discord of a bipartite 
system can be witnessed by accessing only one of the two subsystems \cite{GB}. 
The method extends a general theoretical scheme for the detection of 
initial correlations in the dynamics of open quantum systems developed in 
Ref.~\cite{WITNESS}, which has been recently realized experimentally 
\cite{Tang,Smirne11}. Typically, an open quantum system represents a 
well-controlled quantum system which is coupled to a complex, largely
inaccessible environment and therefore constitutes a natural setting in 
which we could benefit from the method described in this paper.

Our strategy for the construction of a local witness for the quantum discord 
in a bipartite system is based on a local dephasing operation, describing 
measurements carried out on one of the subsystems, which leaves the marginal 
states invariant while erasing all quantum discord between the two subsystems. 
When the subsequent time evolution of the composite, bipartite system is changed 
by this dephasing operation, one can conclude that the original state has a 
non-vanishing quantum discord. A suitable local witness for quantum discord is 
thus given by any appropriate measure for the distance between the time-evolved 
reduced subsystem states obtained from the total system states corresponding to
the evolution with and without local dephasing operation \cite{GB}.

In the present paper we develop the details of this method and provide a study of 
its applications to thermal equilibrium states of generic complex quantum systems. 
In order to assess the performance of our witness for quantum discord we 
compare the actual dynamics under randomly generated Hamiltonians with the 
mean values and fluctuations obtained from the average over the unitary group 
equipped with the Haar measure, employing results of Ref.~\cite{Unitaries}. We 
conclude with the formulation of a general ergodicity-type hypothesis which
relates the average of the local witness over the unitary group to the time average
of the witness obtained for a generic system dynamics.

\section{Quantum discord and local dephasing operation}
Throughout this paper we deal with a bipartite Hilbert space 
$\mathcal{H}=\mathcal{H}_A\otimes\mathcal{H}_B$, composed of local Hilbert 
spaces $\mathcal{H}_A$ and $\mathcal{H}_B$ with dimensions $d_{A}$ and 
$d_{B}$, respectively. A state $\rho$ of the composite system has zero discord 
with respect to subsystem $A$ if and only if it can be written as 
\cite{MODIREVIEW}
\begin{align}\label{eq.defzerod}
\rho=\sum_ip_i|i\rangle\langle i|\otimes\rho_B^i,
\end{align}
with a basis $\{|i\rangle\}$ of $\mathcal{H}_A$, a probability distribution $\{p_i\}$, 
and a set of arbitrary quantum states $\{\rho_B^i\}$. States of zero discord are 
considered as classical. In the following we use this asymmetric 
definition, expressing classicality with respect to subsystem $A$. The reduced 
density operator $\rho_A=\sum_ip_i|i\rangle\langle i|$ is obtained from $\rho$ via 
the partial trace over subsystem $B$. We introduce the quantum operation 
\begin{equation} \label{PHI}
 \Phi(X_A)=\sum_i|i\rangle\langle i|X_A|i\rangle\langle i|
\end{equation}
which represents a completely positive and trace preserving linear map acting
on operators $X_A$ of subsystem $A$. The definition in 
Eq.~(\ref{eq.defzerod}) is then equivalent to 
the following statement: A state $\rho$ has zero discord if and only if the operation
\begin{align}\label{eq.ld}
(\Phi\otimes\mathbb{I}_B)\rho=\sum_i\Pi_i\rho\Pi_i
\end{align}
leaves the state invariant \cite{LUO}, where we have introduced the local projectors 
$\Pi_i=|i\rangle\langle i|\otimes\mathbb{I}_B$ onto the eigenbasis of $\rho_A$, and 
$\mathbb{I}_B$ denotes the identity operation on subsystem $B$. Equation 
(\ref{eq.ld}) defines the local dephasing operation in the eigenbasis of $\rho_A$. 
This operation constitutes the central element for the local detection scheme and 
has a series of important properties \cite{GB}: 
\begin{enumerate}
\item[(i)] The operation (\ref{eq.ld}) can be interpreted as a 
nonselective measurement in the eigenbasis of $\rho_A$, which is fully accessible 
from $\rho$ by measurements in the local subsystem $A$. 
\item[(ii)] None of the two 
reduced density operators $\rho_A$ and $\rho_B$ is affected by 
application of the local dephasing. 
\item[(iii)] The state produced by the local dephasing 
operation is always classical.
\end{enumerate}

Property (i) is easily confirmed: Assume that the reduced state $\rho_A$ 
has been obtained by state tomography. After diagonalization this 
yields the local eigenbasis $\{|i\rangle\}$. The nonselective 
measurement in this basis is described by the operation (\ref{PHI}), which by 
extension to the total Hilbert space results in the local dephasing operation 
(\ref{eq.ld}) associated with the state $\rho$. Furthermore, this operation 
describes complete decoherence in the basis $\{|i\rangle\}$: The diagonal 
elements of any operator represented in this basis are left unchanged while all 
off-diagonal terms are set to zero.

To prove property (ii) we write the total state as
$\rho=\sum_{\alpha}R_A^{\alpha}\otimes R_B^{\alpha}$, where
$R_A^{\alpha}$ and $R_B^{\alpha}$ are operators on $\mathcal{H}_A$ and 
$\mathcal{H}_B$, respectively. The state after 
application of the local dephasing operation will be denoted by $\rho'=(\Phi\otimes
\mathbb{I}_B)\rho$. Its corresponding reduced density operator $\rho'_B$ of 
subsystem $B$ will be unchanged, since only the identity operation is applied to 
this part of the Hilbert space:
\begin{align}
\rho'_B=\text{Tr}_{A}\rho'&=\text{Tr}_{A}\sum_{\alpha}\Phi(R_A^{\alpha})\otimes 
R_B^{\alpha}\notag\\
&=\sum_{\alpha}\text{Tr}\left\{\Phi(R_A^{\alpha})\right\}R_B^{\alpha}\notag\\
&=\sum_{\alpha}\text{Tr}\left\{R_A^{\alpha}\right\}R_B^{\alpha}=\text{Tr}_{A}\rho=
\rho_B.
\end{align}
The reduced state of subsystem $A$ is not altered since the measurement is 
performed in its own eigenbasis:
\begin{align}
 \rho'_A=\text{Tr}_{B}\rho'&=\text{Tr}_{B}\sum_{\alpha}\sum_i|i\rangle\langle i|
 R_A^{\alpha}|i\rangle\langle i|\otimes R_B^{\alpha}\notag\\
&=\sum_{\alpha}\sum_i|i\rangle\langle i|\text{Tr}\left\{R_B^{\alpha}\right\}\langle i|
R_A^{\alpha}|i\rangle\notag\\
&=\sum_{i}p_i\ket{i}\!\bra{i}=\rho_A,
\end{align}
where $p_i=\sum_{\alpha}\text{Tr}\left\{R_B^{\alpha}\right\}\langle i|
R_A^{\alpha}|i\rangle=\langle i|\rho_A|i\rangle$.

Finally, property (iii) is obvious since $\rho'$ can be readily cast into 
the form of Eq.~(\ref{eq.defzerod}) with $p_i\rho_B^i=\sum_{\alpha}\langle i|
R_A^{\alpha}|i\rangle R_B^{\alpha}$. The combination of all three properties 
leads to an additional interpretation: Performing a 
nonselective measurement in the local eigenbasis, i.~e., applying the 
corresponding local dephasing operation to a state $\rho$ erases the quantum 
discord in $\rho$ while leaving its marginals unchanged.

We end this section with remarks on two special situations. First, in the case of 
degeneracies in the spectrum of $\rho_A$, the local basis $\{|i\rangle\}$ in
Eq.~(\ref{eq.defzerod}) is not uniquely determined by the local state $\rho_A$.
Performing a local dephasing operation in an arbitrary eigenbasis of $\rho_A$
can then change the given total state $\rho$ even if it has zero discord. 
However, in the following we may ignore the possibility of degenerate 
local states since they form a set of zero measure.
Second, if the local state tomography yields a pure state $\rho_A=|\varphi\rangle
\langle\varphi|$, no further action is required. It is already safe to conclude that no 
total correlations exist between the two subsystems and the total state is a product 
state, $\rho=|\varphi\rangle\langle\varphi|\otimes\rho_B$. Specifically, this situation 
is encountered in experiments if one of the subsystems is prepared in a pure state. 
Total correlations in terms of the distance to the corresponding product state can 
be witnessed on the basis of an arbitrary local operation using the 
contraction property of the trace distance \cite{WITNESS}.

\section{Local witness for quantum discord}

For simplicity, we assume that the composition of the systems $A$ and $B$ forms 
a closed system. We will see below that this assumption can be 
dropped. The dynamics of a closed system is described by a unitary time evolution 
operator $U_t$, propagating states from time $0$ to time $t$. Tracing 
over subsystem $B$ yields the reduced density matrix at time $t$, 
$\rho_A(t)=\text{Tr}_B\{U_t\rho U^{\dagger}_t\}$. 
The local detection method is based on the following idea: First, the accessible 
part of the unknown initial state $\rho$ is measured, yielding the state $\rho_A$ 
and its eigenbasis. After the reference state $\rho'$ is produced by local 
nonselective measurement of the total state in this basis, we compare the 
dynamics of the two reduced states $\rho_A(t)$ and 
$\rho'_A(t)=\text{Tr}_B\{U_t\rho'U^{\dagger}_t\}$. The difference of these 
states can be quantified by an arbitrary operator distance
\begin{align}\label{eq.dist}
 \text{dist}(t)=\|\rho_A(t)-\rho'_A(t)\|^2
 =\|\text{Tr}_B\{U_t(\rho-\rho')U_t^{\dagger}\}\|^2.
\end{align}
First, note that $\text{dist}(0)=0$ due to property (ii) of the local dephasing map. 
On the other hand, if we find an instant of time $t>0$ for which 
$\text{dist}(t)>0$, we can conclude that $\rho$ and $\rho'$ must be different states. 
This in turn implies that $\rho$ has nonzero discord which enables us 
to locally witness bipartite quantum discord \cite{GB}.

Since the states $\rho$ and $\rho'$ differ only in their quantum discord, a possible 
measure for the amount of discord is given by the distance \cite{LUO}
\begin{align}\label{eq.disc}
\mathcal{D}(\rho)=\|\rho-\rho'\|^2.
\end{align}
Until this point all results are independent of the specific choice of distance. For later applications we 
choose the squared Hilbert-Schmidt norm $\|A\|^2=\text{Tr}A^{\dagger}A$, which has also been used in a similar context under the term geometric measure for quantum discord \cite{LuoFu,DVB}. With 
this choice, Eq.~(\ref{eq.disc}) can be written as a difference of purities \cite{GB}. 
More generally, for any map of the form 
$\Phi(X_A)=\sum_i\pi_iX_A\pi_i$  with a complete set of mutually orthogonal 
projection operators $\pi_i$ we have:
\begin{align}
  \left\|\rho-\left(\Phi\otimes\mathbb{I}_B\right)\rho\right\|^2 & = 
  \text{Tr}\left\{\rho^2\right\}-\text{Tr}\left\{\left[\left(\Phi\otimes\mathbb{I}_B\right) 
  \rho\right]^2\right\}\notag\\
 & = \mathcal{P}\left(\rho\right)-\mathcal{P}\left(\left(\Phi\otimes\mathbb{I}_B\right)
 \rho\right), \label{eq.hsequalspuritydifference}
\end{align}
with the purity $\mathcal{P}(\rho)=\text{Tr}\{\rho^2\}$. To prove this 
relation we write the left-hand side of this equation as
\begin{align}
&\left\|\rho-\left(\Phi\otimes\mathbb{I}_B\right)\rho\right\|^2\notag\\
&=\mathcal{P}(\rho)-2\,\text{Tr}\{\rho(\Phi\otimes\mathbb{I}_B)\rho\}+\mathcal{P}\left((\Phi\otimes\mathbb{I}_B)\rho\right).
\label{eq.hsreferencestate}
\end{align}
Making use of the Kraus representation of $\Phi$, we obtain
\begin{align}
\mathcal{P}\left((\Phi\otimes\mathbb{I}_B)\rho\right)&=
\text{Tr}\left\{\sum_{\alpha,\beta,i,j}\delta_{ij}\pi_iR_A^{\alpha}\pi_iR_A^{\beta}
\pi_j\otimes R_B^{\alpha} R_B^{\beta}\right\}\notag\\
&=\sum_{\alpha,\beta,i}\text{Tr}\left\{R_A^{\alpha}\pi_i R_A^{\beta}\pi_i\right\}
\text{Tr}\left\{R_B^{\alpha} R_B^{\beta}\right\}\notag\\
&=\text{Tr}\left\{\sum_{\alpha,\beta,i}R_A^{\alpha}\pi_i R_A^{\beta}\pi_i\otimes
R_B^{\alpha} R_B^{\beta}\right\}\notag\\
&=\text{Tr}\left\{\rho\left(\Phi\otimes\mathbb{I}_B\right)\rho\right\},
\end{align}
which proves Eq.~(\ref{eq.hsequalspuritydifference}). Moreover, this adds a nice 
operational interpretation to the measure $\mathcal{D}(\rho)$ in terms of the 
purity-decreasing effect of the local dephasing operation.
Note that if the state $\rho$ is pure, the expression $\mathcal{D}(\rho)$ yields the 
generalized concurrence \cite{GB}, a well-known entanglement measure 
\cite{WOOTERS,RUNGTA}, illustrating the equivalence of discord and 
entanglement in the case of pure states.

For distance measures which are contractive under the action of trace-preserving 
quantum operations, $\text{dist}(t)$ provides a lower bound for the quantum 
discord expressed by $\mathcal{D}(\rho)$. Using for example the trace 
norm defined by $\|A\|_1=\text{Tr}\sqrt{A^{\dagger}A}$ we obtain:
\begin{align}
 \mathcal{D}(\rho) \geq \text{dist}(t).
\end{align}
Even though the Hilbert-Schmidt distance is not contractive under 
trace-preserving operations, one can derive a lower bound for the 
quantum discord in terms of $\text{dist}(t)$, employing the contractivity of the trace 
distance and well-known upper and lower bounds for the Hilbert-Schmidt 
distance in terms of the trace distance:
\begin{align}
\mathcal{D}(\rho)\geq\frac{1}{d_Ad_B}\text{dist}(t).
\end{align}

We note that this method can even be extended to general linear 
time-evolutions given by a family of quantum dynamical maps 
$\Lambda_t$, such that
$\rho(t)=\Lambda_t(\rho)$ and $\rho'(t)=\Lambda_t(\rho')$, which yields
\begin{align}
\text{dist}(t)=\|\text{Tr}_B\{\Lambda_t(\rho-\rho')\}\|^2.
\end{align}
Thereby the scheme can be used to detect correlations also in bipartite systems 
under additional dissipation caused by the coupling to an external 
environment. For the rest of this paper, we will restrict to the case of unitary 
evolution.

\section{Performance of the witness and examples}

The above method may fail to detect correlations depending on the time evolution 
$U_t$. Consider for instance the trivial case of two uncoupled subsystems. The 
time evolution factorizes, $U=U_A\otimes U_B$, where we omit the time 
argument. In this case, no signature of the total state will be visible in the reduced 
system dynamics, which can be seen easily by decomposing 
$\rho-\rho'=\sum_{\alpha}D_A^{\alpha}\otimes D_B^{\alpha}$:
\begin{align}
\text{Tr}_B\left\{U(\rho-\rho')U^{\dagger}\right\}&=\text{Tr}_B\left\{\sum_{\alpha}
U_AD_A^{\alpha}U_A^{\dagger}\otimes U_BD_B^{\alpha}U_B^{\dagger}\right\}
\notag\\
&=\sum_{\alpha}U_AD_A^{\alpha}U_A^{\dagger}\text{Tr}\left\{D_B^{\alpha}\right\}
\notag\\
&=U_A\text{Tr}_B\left\{\rho-\rho'\right\}U_A^{\dagger}=0.
\end{align}
The question is thus, what is the performance of the method for 
generic systems? In order to answer this question we make use of a recently 
developed approach based on unitary average values \cite{GB,Unitaries}. In order 
to obtain an estimate for the quantity $\text{dist}(t)$, we replace $U_t$ with a 
random unitary matrix $U$ and determine the average integrating over the uniform 
Haar measure $d\mu$. According to ensemble theory, the average value is 
expected to reflect the behavior of generic complex quantum systems. We denote 
unitary average values by angular brackets,
\begin{align}
\left\langle F(U)\right\rangle=\int d\mu(U)F(U).
\end{align}
The Hilbert-Schmidt distance for an arbitrary pair of states $\rho$ and $\rho'$ 
yields the average value \cite{GB}
\begin{align}
\mu \equiv
\left\langle\left\|\text{Tr}_B\left\{U(\rho-\rho')U^{\dagger}\right\}\right\|^2
\right\rangle=\frac{d_A^2d_B-d_B}{d_A^2d_B^2-1}\left\|\rho-\rho'\right\|^2,
\label{eq.uniformtwostate}
\end{align}
and the variance \cite{Unitaries}
\begin{eqnarray}
 s^2 &\equiv&
 \text{Var}(\left\|\text{Tr}_B\left\{U(\rho-\rho')U^{\dagger}\right\}\right\|^2)
 \nonumber \\
 &=& c_1 (\text{Tr}\{(\rho-\rho')^2\})^2 + c_2\text{Tr}\{(\rho-\rho')^4\},
 \label{eq.twostatevariance}
\end{eqnarray}
with the coefficients $c_1$ and $c_2$ given by
\begin{align} \label{C1-C2}
c_1&=\frac{2(15-4d_A^2d_B^2+d_A^4d_B^4)(d_A^2-1)(d_B^2-1)}{(36-13d_A^2d_B^2+d_A^4d_B^4)(d_A^2d_B^2-1)^{2}},\notag\\
c_2&=\frac{-10d_Ad_B(d_B^2-1)(d_A^2-1)}{d_A^2d_B^2(d_A^2d_B^2-7)^2-36}.
\end{align}
Inserting $\rho'=(\Phi\otimes\mathbb{I}_B)\rho$ into Eq.~(\ref{eq.uniformtwostate}), 
we find that the average increase of the local distance is directly proportional to 
the squared Hilbert-Schmidt distance of the original state $\rho$ to its locally 
dephased reference state $\rho'$, which we had previously defined as 
$\mathcal{D}(\rho)$, a measure for quantum discord. This result also holds for a 
more general average, which is performed only over the eigenvectors of the 
Hamiltonian while the time dependence and the eigenvalue distribution are 
retained, see Refs.~\cite{GB,Unitaries}. 

From Eqs.~(\ref{eq.uniformtwostate})-(\ref{C1-C2}) we find that for large
$d_B$ the relative
fluctuations are given by
\begin{equation}
 \frac{s}{\mu} \approx \sqrt{\frac{2}{d_A^2-1}}.
\end{equation}
This ratio is always smaller than one and decreases as $s/\mu\sim 1/d_A$ for 
large $d_A$. Thus we see that the standard deviation $s$ is 
at most of the same order of magnitude as the mean value $\mu$ \cite{Unitaries}. 
This statement is confirmed by the numerical studies discussed below. 
Since the median ($50\%$-quantile) of a random number always 
lies in the range $\mu\pm s$, we find that the squared reduced system 
Hilbert-Schmidt distance is larger than $\mu-s$ with a probability of at least 
$50\%$. We conclude that for generic systems the quantum discord in the initial 
state will be successfully detected by the present method with high probability.

The main purpose of the unitary average value is to demonstrate the general 
reliability of the presented method. However, we note that from 
Eq.~(\ref{eq.uniformtwostate}) we see that if the unitary average of the local 
distance could be measured, it could be used not only to witness the discord in the 
initial states, but also to quantify it. Even though the number of gates needed for 
the realization of Haar-random unitary operators scales exponentially with the 
number of qubits involved, there have been efforts aiming at the realization of 
unitary averages with methods which are experimentally feasible 
\cite{Emerson1,Emerson2,Dankert}.

\subsection{Simple example of pure states}

\begin{figure}
 \centering
\includegraphics[width=.45\textwidth]{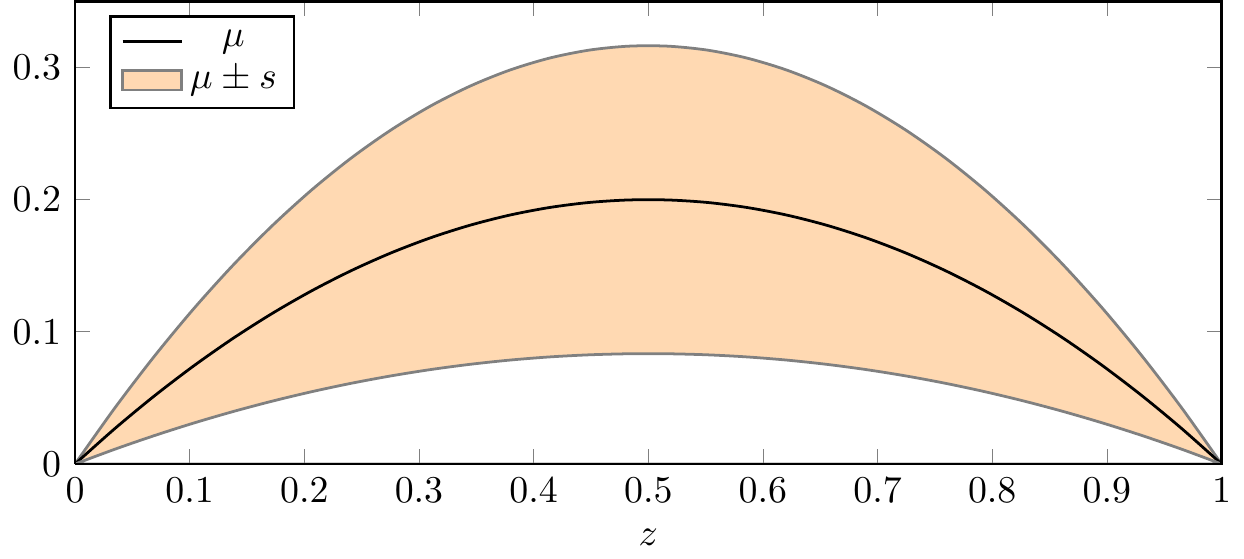}
\caption{(Color online) The plot shows the dependence of the unitary average 
value $\mu$ and the variance $s^2$ on the parameter $z$ for $\rho_z$. 
The relative error is constant at $s/\mu\:{\approx0.58}$.}
\label{fig.wernerLDM}
\end{figure}

As a first simple illustration of this method, we consider pure states $\rho_z=|\Psi_z\rangle\langle\Psi_z|$, with
\begin{align}
|\Psi_z\rangle=\sqrt{z}|00\rangle+\sqrt{1-z}|11\rangle,
\end{align}
and $0 \leq z \leq 1$. The reduced system state is given by $\rho_A=z|0\rangle\langle0|+(1-z)|1\rangle\langle1|$, 
and to produce the reference state by local dephasing we project onto the 
operators $|0\rangle\langle0|$ and $|1\rangle\langle1|$,
\begin{align}
 \rho'_z=(\Phi\otimes\mathbb{I}_B)\rho_z=\sum_{i=0,1}
 (|i\rangle\langle i|\otimes\mathbb{I}_B)\rho_z(|i\rangle\langle i|\otimes\mathbb{I}_B).
\end{align}
Thus, we obtain the reference state
\begin{align}
 (\Phi\otimes\mathbb{I}_B)\rho_z=\:&z\ket{00}\!\bra{00}+(1-z)\ket{11}\!\bra{11}.
\end{align}
The mixedness of the reduced state of $\rho_z$ stems from the entanglement in $|\Psi_z\rangle$. On the other hand, $\rho'_z$ is only classically correlated but passes its own mixedness on to the reduced state. Since both states yield the same reduced density matrix, the nature of the total state cannot be revealed on the basis of the reduced system at the initial time. However, if the subsequent time evolution in the subsystem is taken into account, it is possible to distinguish between the total states with and without quantum correlations.

The generic increase of the distance in the reduced system is given by Eq.~(\ref{eq.uniformtwostate}), which leads to
\begin{align}
 \mu =
 \left\langle\left\|\text{Tr}_B\left\{U(\rho_z-\rho'_z)
 U^{\dagger}\right\}\right\|^2\right\rangle=\frac{2}{5}\mathcal{D}(\rho_z),
\end{align}
where $\mathcal{D}(\rho_z)=2(1-z) z$ is proportional to the square of the concurrence.

The variance is given by Eq.~(\ref{eq.twostatevariance}), which for this state yields
\begin{align}
  s^2=\text{Var}\left(\left\|\text{Tr}_B\left\{U(\rho_z-\rho'_z) U^{\dagger}\right\}\right\|^2\right)=\frac{38}{175} (z-1)^2 z^2.
\end{align}

The relative error is constant for all values of $z$ and amounts to 
$s/\mu=\sqrt{19/56}\:{\approx0.58}$. The relatively large value of the variance is 
explained by the low dimensions of system and environment. A plot showing the 
dependence of expectation value and variance on the parameter $z$ is given in 
Fig.~\ref{fig.wernerLDM}.

\subsection{Random Gibbs states of $2\times d_B$ systems}

\begin{figure}
\centering
\includegraphics[width=.45\textwidth]{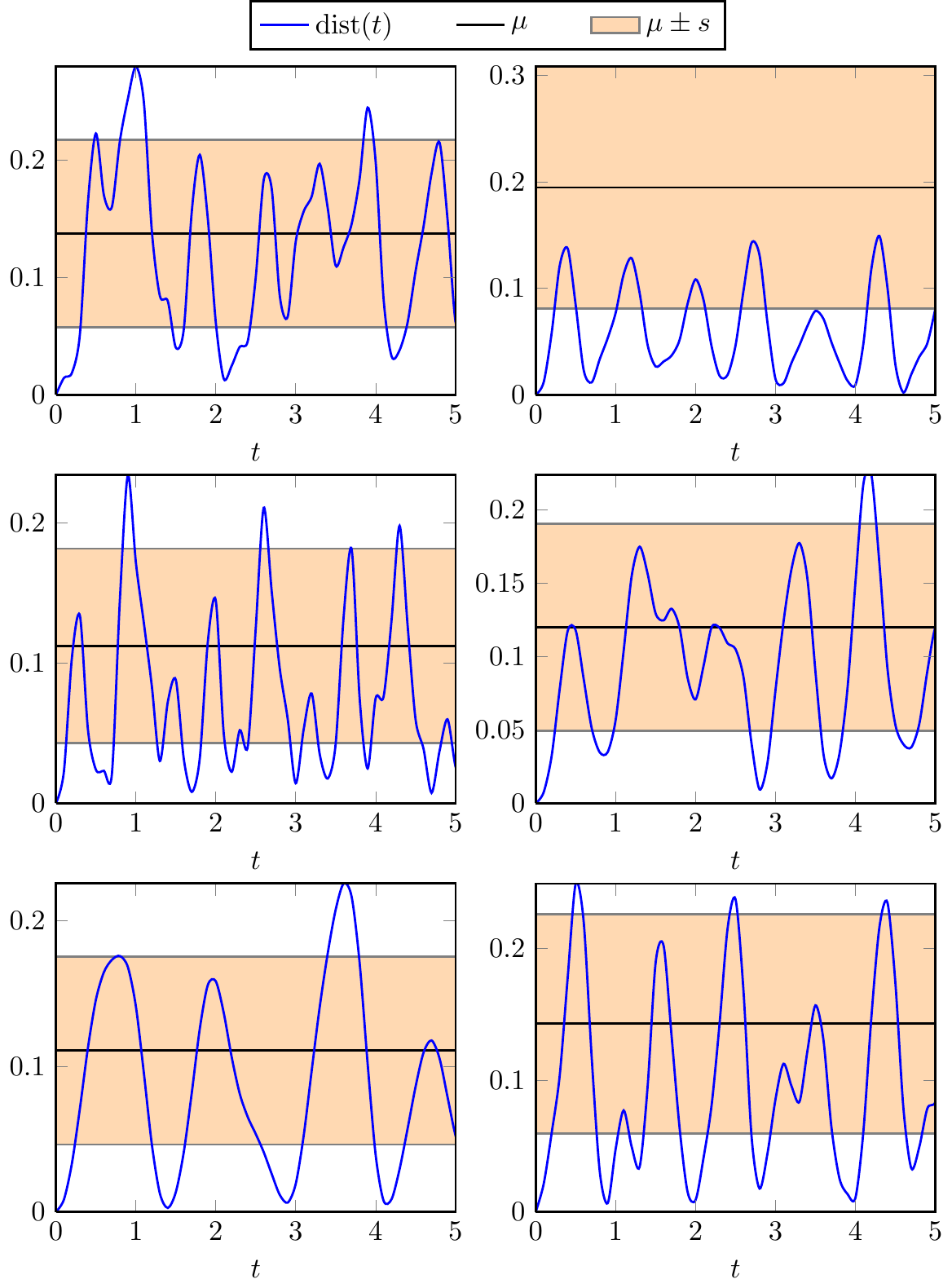}
\caption{(Color online) Comparison of the unitary average with the actual time 
evolution for the Gibbs states of six randomly picked two-qubit Hamiltonians 
($d_A=d_B=2$) at fixed temperature $\beta=1$. The pictures show the value of 
the Hilbert-Schmidt distance after applying the local detection method to the Gibbs 
state.}
\label{fig.randomHtwoqubits}
\end{figure}

\begin{figure}[tb]
\centering
\includegraphics[width=.48\textwidth]{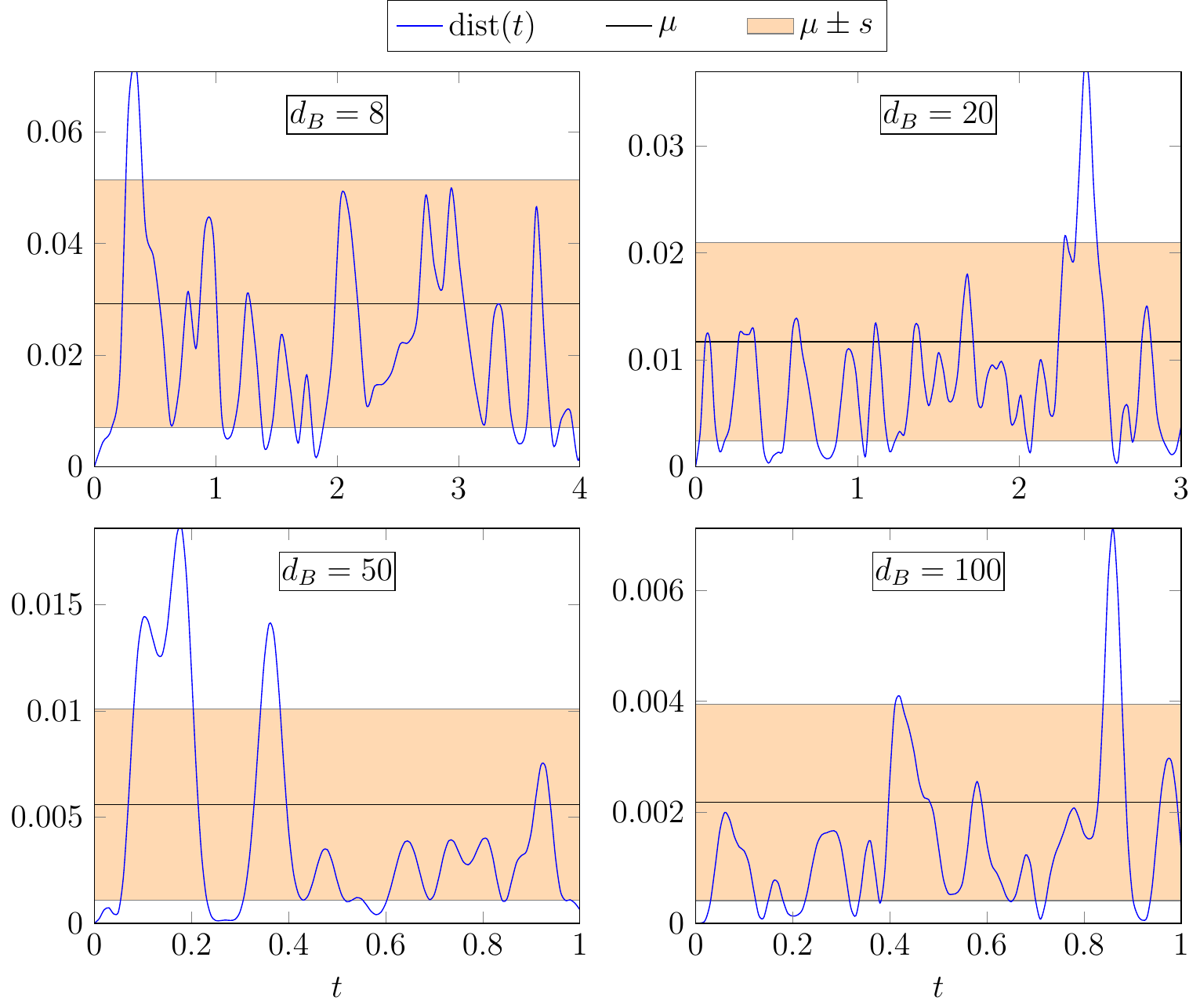}
\caption{(Color online) Comparison of the unitary average with the actual time 
evolution for the Gibbs states of four randomly picked Hamiltonians for a qubit 
coupled to environments with different dimensions at fixed temperature 
$\beta=1$.}
\label{fig.randomHseveralds}
\end{figure}

In this section we demonstrate the local detection scheme for Gibbs 
states of randomly generated $d$-dimensional Hamiltonians. Once such a random $H$ has been generated \cite{MEZZADRI2007}, 
the Gibbs state can easily be obtained as $\rho_G=e^{-\beta H}/Z$, with the 
partition function $Z=\text{Tr}e^{-\beta H}$, $\beta=1/kT$, temperature $T$, and 
the Boltzmann constant $k$. We consider the total Hilbert space to be 
$2d_B$-dimensional, i.~e., the system Hilbert space $\mathcal{H}_A$ is 
two-dimensional. Employing the product basis $\{\ket{0},\ket{1}\}
\otimes\{\ket{\chi_i}\}_{i=1}^{d_B}$, where $\{\ket{\chi_i}\}$ denotes an arbitrary 
fixed basis of $\mathcal{H}_B$, the Gibbs state $\rho_G$ can be written as
\begin{align}
 \rho_G=&\:\sum_{i,j}a^{00}_{ij}\ket{0}\!\bra{0}\otimes\ket{\chi_i}\!\bra{\chi_j}
 +\sum_{i,j}a^{01}_{ij}\ket{0}\!\bra{1}\otimes\ket{\chi_i}\!\bra{\chi_j}\notag\\
&+\:\sum_{i,j}a^{10}_{ij}\ket{1}\!\bra{0}\otimes\ket{\chi_i}\!\bra{\chi_j}+\sum_{i,j}
a^{11}_{ij}\ket{1}\!\bra{1}\otimes\ket{\chi_i}\!\bra{\chi_j}.
\end{align} 
Hence, the reduced density operator of subsystem $A$ can be 
represented by the matrix
\begin{align}
 \rho_A=\text{Tr}_B\rho_G=\begin{pmatrix}
    \sum_ia^{00}_{ii}&\sum_ia^{01}_{ii}\\
    \sum_ia^{10}_{ii}&\sum_ia^{11}_{ii}
   \end{pmatrix}.
\end{align}
On the basis of the eigenvectors $\{\ket{\widetilde{0}},\ket{\widetilde{1}}\}$ of this 
$(2\times2)$-matrix, the local dephasing map is expressed as
\begin{align}
 (\Phi\otimes\mathbb{I}_B)\rho=\Pi_{\widetilde{0}}\rho\Pi_{\widetilde{0}}
 +\Pi_{\widetilde{1}}\rho\Pi_{\widetilde{1}},
\end{align}
with $\Pi_{\widetilde{i}}=\ket{\widetilde{i}}\!\bra{\widetilde{i}}\otimes\mathbb{I}_B$. 
Application of this map to the original Gibbs state $\rho_G$ creates the reference 
state $(\Phi\otimes\mathbb{I}_B)\rho_G$. Next, we examine the dynamics of the 
distance of the two reduced system states by creating the corresponding time 
evolution operator $U_t=\exp\{-i Ht\}$ from the same randomly generated 
Hamiltonian $H$. 
The distance is given as a function of $t$ by:
\begin{align}
 \text{dist}(t)=\|\text{Tr}_B\{U_t(\rho_G-(\Phi\otimes\mathbb{I}_B)\rho_G)
 U_t^{\dagger}\}\|^2.
\end{align}
On the other hand we can obtain the unitary expectation value and its variance for 
the same quantity by Eqs.~(\ref{eq.uniformtwostate}) and 
(\ref{eq.twostatevariance}), which in this case yield
\begin{align}\label{eq.mu}
 \mu&=\left\langle\left\|\text{Tr}_B\left\{U(\rho_G-(\Phi\otimes\mathbb{I}_B)\rho_G)
 U^{\dagger}\right\}\right\|^2\right\rangle\notag\\&=\frac{3d_B}{4d_B^2-1}\left\|
 \rho_G-(\Phi\otimes\mathbb{I}_B)\rho_G\right\|^2
\end{align}
and
\begin{align}\label{eq.s}
 s^2=&\:\text{Var}\left(\left\|\text{Tr}_B\left\{U(\rho_G-(\Phi\otimes\mathbb{I}_B)
 \rho_G)U^{\dagger}\right\}\right\|^2\right)\notag\\=&\:\frac{3 (15 - 16 d_B^2 + 16
  d_B^4)}{2 (1 - 4 d_B^2)^2 (4 d_B^2-9)}\left\|\rho_G-(\Phi\otimes\mathbb{I}_B)
  \rho_G\right\|^4\notag\\&-\:\frac{15 d_B}{9 - 40 d_B^2 + 16 d_B^4}\text{Tr}
  \left\{(\rho_G-(\Phi\otimes\mathbb{I}_B)\rho_G)^4\right\}.
\end{align}

\begin{figure}[tb]
\centering
\includegraphics[width=.45\textwidth]{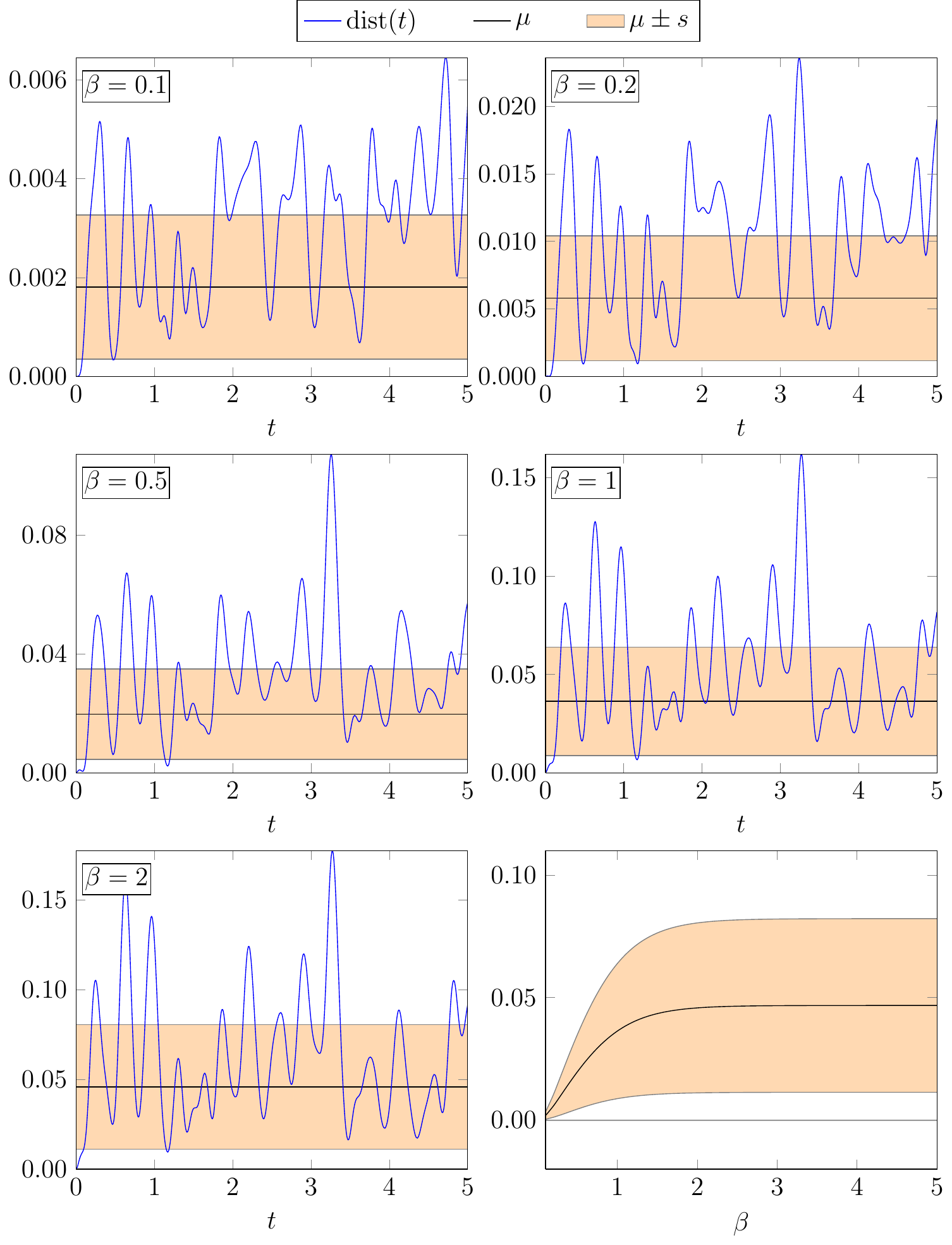}
\caption{(Color online) Dependence of the discord of a randomly picked fixed 
thermal state on the temperature for $d_B=8$. For higher temperatures (lower 
$\beta$), discord diminishes. The picture on the lower right shows the average 
value and one standard deviation as a function of the inverse temperature 
$\beta$.} \label{fig.randomHseveralTs}
\end{figure}

We have carried out an extensive numerical study of various cases 
with many different parameter sets and initial states.
In the following we present a selection of our results to illustrate the main
features. Figure \ref{fig.randomHtwoqubits} shows a series of time evolutions 
including the corresponding unitary average value $\mu$ and the first 
standard deviation $s$ for six randomly generated $2\times2$ 
Hamiltonians at fixed temperature $\beta=1$. The dependence on the 
environmental dimension is plotted in Fig.~\ref{fig.randomHseveralds}, while 
Fig.~\ref{fig.randomHseveralTs} displays the role of the inverse temperature 
$\beta$. From these simulations we can make a 
number of observations. First, the numerical analysis suggests that generic Gibbs 
states contain quantum 
discord since the function $\text{dist}(t)$ assumes nonzero values for all 
realizations, confirming measure-theoretic studies on the abundance of quantum 
discord \cite{Ferraro,TURKU}. Second, for most of the examples the time evolution 
fits nicely into the margin given by the unitary average within one standard 
deviation, indicated by the highlighted areas. It is of course 
no surprise to find some deviating realizations as in the top right picture of 
Fig.~\ref{fig.randomHtwoqubits}. Third, as becomes obvious by comparison of 
Figs.~\ref{fig.randomHtwoqubits} and \ref{fig.randomHseveralds}, the unitary 
average value depends stronger on the dimensions of system and environment 
than on the actual Hamiltonian. The values in Fig.~\ref{fig.randomHtwoqubits} 
differ only very little between the considered random examples, while in 
Fig.~\ref{fig.randomHseveralds} we see that the average value $\mu$ 
and the standard deviation $s$ decrease significantly with increasing 
environmental dimension. This is mainly caused by the 
dimension-dependent factors in Eqs.~(\ref{eq.mu}) and (\ref{eq.s}). 

Figure \ref{fig.randomHseveralTs} shows how the witness 
$\text{dist}(t)$ changes for different temperatures. We see however that
the overall functional shape remains similar which is due to the fact that the
Hamiltonian is the same in all plots. The bottom right picture shows the 
asymptotic convergence of mean value and variance for decreasing temperature. 
In the high-temperature limit ($\beta\rightarrow0$) the unitary average value, and 
with it the generic effect of the initial correlations on the reduced system vanishes 
as expected since the state becomes closer to a complete mixture, which is a 
state of zero discord. Note that correlations in the low-temperature limit 
of the Gibbs state can be used to reveal the structure of the ground state 
\cite{Smirne10}, which in turn can be associated with a quantum phase transition 
\cite{Sachdev,Osterloh,Vidal,Wu}.

To conclude this section, we recall that a state of nonzero discord cannot be a 
factorized product state \cite{LiLuo08}. On the other hand, factorizing initial 
conditions are commonly assumed in the derivation of master equations for the 
dynamical description of open systems in terms of completely positive maps, see, 
e.g., Refs.~\cite{BREUERBOOK,LINDBLAD1996,WITNESS} and references 
therein. Hence, the present method can also be used to detect deviations from this 
assumption \cite{GB}. Obviously, if the witness is nonzero, a dynamical map which 
is independent of the correlations does not exist. A study of the role of the total 
initial correlations in thermal equilibrium states is presented in 
Ref.~\cite{Smirne10}.

\subsection{An ergodicity-like relation}
The foregoing study shows that unitary averages provide
important and useful information about the time evolution, which may be 
experimentally observable. It was pointed out in Ref.~\cite{Unitaries} that the 
dimension $d_B$ appearing in expressions for the averages must be 
chosen carefully. Formally, it is always possible to artificially increase the 
dimension of the Hilbert space by including an additional Hilbert space which is not 
coupled to the original system. Correspondingly,
the dimension appearing in the expectation value must be regarded 
as an effective dimension, indicating the dimension of the subspace of the Hilbert 
space which actually affects the local dynamics. In general, a suitable, 
effective dimension $d_B^{\mathrm{eff}}$ may be defined via the equality
\begin{align}
&\left\langle\left\|\text{Tr}_B\left\{U(\rho-\rho')U^{\dagger}\right\}\right\|^2
\right\rangle_{\mathrm{eff}}
\notag\\&=\lim_{T\rightarrow\infty}\frac{1}{T}\int\limits_0^Tdt\left\|
\text{Tr}_B\left\{U_t(\rho-\rho')U_t^{\dagger}\right\}\right\|^2.
\end{align}
Thus, we are led to an
ergodicity-like hypothesis for complex generic systems expressing the equivalence 
of the unitary average value and the time average according to the given, actual 
Hamiltonian: For complex generic systems, the effective dimension coincides with 
the dimension of the Hilbert space. The effective dimensions of non-generic 
systems depend not only on the system parameters but also on the observable in 
question. For example, in a partly chaotic system with regular areas, some initial 
states may explore large parts of the state space in the course of their time 
evolution while for different initial conditions only a very limited fraction may be 
visited. The estimation of the dimension of quantum systems is a topic of growing 
interest \cite{dim}.

\section{Conclusion}
The method discussed in this paper allows for the detection of 
quantum discord in bipartite systems when access to only one of the subsystems 
is possible. This situation emerges naturally in the context of open 
quantum systems and quantum communication protocols. The procedure was 
illustrated by application to thermal equilibrium states of random Hamiltonians. 
In order to estimate the performance of the method for generic systems we 
compared the time evolution with the value obtained by averaging over all unitary 
evolutions employing the Haar measure. The mean values 
as well as the fluctuations predicted by the Haar measure were found to be in 
good agreement with the actual time evolution. This fact led to the proposition of 
an ergodicity-like hypothesis, linking unitary average and time average, and to the 
introduction of an effective dimension of the underlying Hilbert space.
Further studies are required, on the one hand to confirm this hypothesis with 
additional examples of generic systems and, on the other hand, 
to obtain the effective dimensions of non-generic systems which 
typically exploit only an effective subspace whose dimension is much lower than 
that of the total Hilbert space.

\acknowledgments
M.G. thanks the German National Academic Foundation for support.

\end{document}